\documentstyle[prl,multicol,aps,epsf]{revtex}
\renewcommand{\widetext}
{\end{multicols}\global\columnwidth42.5pc}
\begin{document}
\newcommand{\be}{\begin{equation}}
\newcommand{\ee}{\end{equation}}
\newcommand{\bea}{\begin{eqnarray}}
\newcommand{\eea}{\end{eqnarray}}
\newcommand{\br}{{\bf r}}
\newcommand{\bk}{{\bf k}}
\newcommand{\bq}{{\bf q}}
\newcommand{\bn}{{\bf n}}
\draft
\title{Wave function correlations on the ballistic scale:\\
Exploring quantum chaos by quantum disorder}
\author{I.~V.~Gornyi$^{1,2,*}$ and A.~D.~Mirlin$^{1,2,\dagger}$}
\address{$^1$Institut
f\"ur Nanotechnologie, Forschungszentrum Karlsruhe, 76021 Karlsruhe,
Germany}
\address{$^2$Institut f\"ur Theorie der Kondensierten Materie,
Universit\"at Karlsruhe, 76128 Karlsruhe, Germany}
\date{\today}
\maketitle
\begin{abstract}
We study the statistics of wave functions in a ballistic chaotic
system. The statistical ensemble is generated by
adding weak smooth disorder. The conjecture of Gaussian
fluctuations of wave functions put forward by Berry and
generalized by Hortikar and Srednicki is proven to hold on
sufficiently short distances, while it is
found to be
strongly violated on
larger scales. This also resolves the conflict between the above
conjecture and the wave function normalization. The method is
further used to study ballistic correlations of wave functions in
a random magnetic field.
\end{abstract}
\begin{multicols}{2}
\narrowtext

{\it Introduction.} Understanding of statistical properties of
eigenfunctions of a quantum system whose classical counterpart is
chaotic and their relation to the underlying classical dynamics
is one of the key problems studied in the field of quantum chaos.
Among various applications, wave function correlations are
important for statistics of electron transport through quantum
dots, see \cite{alhassid,baranger} and references therein. It was
conjectured by Berry \cite{Berry77} that an eigenfunction of a
classically chaotic system (``billiard'') can be represented as a
random superposition of plane waves with fixed absolute value $k$
of the wave vector (determined by the energy $k^2/2m=E$, where
$m$ is the mass and we set $\hbar=1$). This implies Gaussian
statistics of the eigenfunction amplitude $\psi(\br)$,
\begin{equation}
{\cal P} \{ \psi \} \propto
\exp\left[-{\beta \over 2} \! \int d^2\br d^2\br'
\psi^*(\br)C^{-1}(\br,\br')\psi(\br')\right],
\label{Ppsi}
\end{equation}
determined solely by the correlation function
(we consider a two-dimensional system)
\begin{equation}
C(\br_1,\br_2)\equiv\langle\psi^*(\br_1) \psi(\br_2) \rangle=
J_0(k|\br_1-\br_2|)/V.
\label{Crr}
\end{equation}
Here $\beta=1$ or $2$ for a system with preserved (respectively
broken) time reversal symmetry, $V$ is the system area, and
$J_0$ the Bessel function. For definiteness, we will consider the
case $\beta=2$ below; generalization to systems with
$\beta=1$ is straightforward.

Hortikar and Srednicki \cite{Srednicki2} noticed that
Eqs.~(\ref{Ppsi}),\ (\ref{Crr}), which do not depend on any
details of the dynamics, may only be valid for sufficiently small
spatial separation. They generalized Berry's hypothesis and
conjectured the Gaussian statistics (\ref{Ppsi}) with a more
general, system-specific kernel $C(\br_1,\br_2)$ replacing
Eq.~(\ref{Crr}), \begin{equation}
C(\br_1,\br_2)=
\frac{{\rm Im} G_{sc}(\br_1,\br_2)}{\int \! d\br{\rm Im}
G_{sc}(\br,\br)}, \label{sred}
\end{equation}
where $ G_{sc}$ is a semiclassical Green's function
\cite{Srednicki2}. This proposal was supported by the observation
\cite{Srednicki1} that the result obtained
in Ref.~\cite{Prigodin} for two-point correlations in a diffusive
system is consistent with the Gaussian statistics.

The conjecture (\ref{Ppsi}),\ (\ref{sred}), while physically very
appealing, obviously requires a formal derivation. Furthermore,
when taken literally, this conjecture contradicts the
wave function normalization,
\be
\int d\br \left[\langle|\psi^2(\br) \psi^2(\br')| \rangle-
\langle|\psi^2(\br)|\rangle \langle|\psi^2(\br')| \rangle
\right]=0,
\label{norm}
\ee
since the integrand is equal to $C^2(\br,\br')>0$ according to
(\ref{Ppsi}). Therefore, limits of validity of this
conjecture have to be understood. All this points to a need in a
systematic study of wave function statistics in ballistic
systems, which is the aim of the present paper.

{\it Eigenfunction statistics in a ballistic system.}
In order to speak about the wave function statistics
${\cal P} \{ \psi(\br) \}$, one should first define an ensemble
over which the averaging goes. Such an ensemble can be generated
\cite{BMM2} by adding to a system under consideration
a random potential $U(\br)$ characterized by a correlation
function $W(\br-\br')=\langle U(\br)U(\br')\rangle$
with a correlation length $d$.
Parameters of this random potential are assumed to
satisfy $k^{-1} \ll d \ll l_s \ll L \ll l_{tr}$,
where $l_s \ (l_{tr})$ is the single-particle
(respectively transport) mean free path, and $L$ is
the characteristic size of the system. The condition
$l_{tr}\gg L$ ensures that the additional disorder
does not influence the classical dynamics of the system,
while the inequality $l_s \ll L$ guarantees that the ensemble of
quantum systems is large enough to produce meaningful result.
Note that the potential is smooth, $kd \gg 1$,
since $l_{tr}/l_s \sim (kd)^2$. On the technical side,
introducing the additional random potential allows us to
apply, with a proper generalization, methods developed earlier
for diffusive systems (see \cite{PhysRep} for a review).

After the ensemble averaging, the problem is described
by a ballistic $\sigma$-model of a supermatrix field
$Q(\br,\bn)$ with the action \cite{woelfle84,amw,TSE,BMM2}
\begin{eqnarray}
S[Q]&=&{\rm Str}\ln\left[E-{\hat H}+i\eta\Lambda
-{i\over 2}\int \! d\bn' Q(\br,\bn')w(\bn,\bn')\right]\nonumber
\\
&-&\frac{\pi\nu}{4}\int d^2\br d\bn d\bn'  {\rm Str}Q(\br,\bn)
w(\bn,\bn')Q(\br,\bn'),
\label{SQ}
\end{eqnarray}
where ${\hat H}$ is the Hamiltonian of the system
(without disorder), $\nu$ is the density of states,
$w(\bn,\bn')=2\pi\nu W(k|\bn-\bn'|)$ is the scattering
cross-section by the random potential, and $\bn$ is a unit vector
characterizing the direction of velocity on the energy surface.
On the scales $\gg l_s$ Eq.~(\ref{SQ}) reduces to the form
proposed in Refs.~\cite{MK,AASA}.
The two-point correlation function of the wave function
intensities is expressed in this approach as \cite{PhysRep,BMM2}
\begin{eqnarray}
\langle|\psi^2(\br_1) \psi^2(\br_2)|\rangle
&=&\lim_{\eta \to 0}\frac{\eta\Delta}{\pi}
\langle [G_{11}(\br_1,\br_1)G_{22}(\br_2,\br_2)\nonumber \\
&+& G_{12}(\br_1,\br_2)
G_{21}(\br_2,\br_1)] \rangle_{S[Q]},
\label{C2fG}
\end{eqnarray}
where $\Delta$ is the mean level spacing, ${\hat G}$ is
the Green's function in the field $Q$, \begin{equation}
{\hat G}=\left[E-{\hat H}+i\eta\Lambda-{i\over 2}\int d\bn'
Q(\br,\bn')w(\bn,\bn')\right]^{-1},
\label{fG}
\end{equation}
and the subscripts $1,2$ refer to the advanced-retarded
decomposition (the boson-boson components being implied). We
first evaluate Eq.~(\ref{C2fG}) in the zero-mode approximation,
$Q(\br)=Q_0$. The Green's function (\ref{fG}) is given in the
leading order by
\bea
G_0(\br,\br')&=& i {\rm Im} G^R(\br,\br')Q_0+
{\rm Re} G^R(\br,\br'),
\label{G0} \\
G^R(\br,\br')&=&\langle \br | (E-{\hat H} +
i/{2\tau_s})^{-1} |\br'\rangle.
\label{GRA}
\eea
If the points $\br,\br'$ are separated by a distance $\gg l_s$
from the billiard boundary, Eq.~(\ref{GRA}) reduces to
$G^R(\br,\br')=J_0(k|\br-\br'|)e^{-|\br-\br'|/2l_s}.$
Substituting Eq.~(\ref{G0}) in (\ref{C2fG}) and expanding the
action (\ref{SQ}) up to the linear-in-$\eta$ term,
$S[Q]\simeq
\pi\nu\eta V {\rm Str} Q_0 \Lambda $, one finds, in
full analogy with the case of diffusive systems,
\bea
&&V^2\langle|\psi^2(\br_1) \psi^2(\br_2)|\rangle
\simeq 1+k_q(\br_1,\br_2); \label{ZM0} \\
&&k_q(\br,\br')={\rm Im}G^R(\br,\br')
{\rm Im}G^R(\br',\br)/(\pi\nu)^2,
\label{kq}
\eea
with the two contributions on the r.h.s. of (\ref{ZM0})
originating from the terms $\langle G_{11}G_{22} \rangle $ and
$\langle G_{12}G_{21}\rangle$ in (\ref{C2fG}), respectively.
The result (\ref{ZM0}), corresponding exactly to the conjecture
(\ref{Ppsi}),\ (\ref{sred}) of the Gaussian statistics, is in
conflict with the wave function normalization, as explained
above.

To resolve this problem, we evaluate the term
$\langle G_{11}G_{22} \rangle $
more accurately by expanding the Green's function
(\ref{fG}) to the order $\eta$ and the action (\ref{SQ}) to the
order $\eta^2$. While these terms (usually neglected in the
$\sigma$-model calculations) are of the next order in $\eta$
and may be naively thought to vanish in the limit $\eta\to 0$
performed in (\ref{C2fG}), this is not so, since $Q_0 \propto
\eta^{-1}$.
As a result, we get in the zero-mode
approximation
\bea
&&V^2 \langle|\psi^2(\br_1) \psi^2(\br_2)|\rangle_{\rm ZM}-1
\nonumber \\
&&= k_q(\br_1,\br_2)-{\bar k}_q(\br_1)-
{\bar k}_q(\br_2)+{\bar{\bar k}}_q
\label{C2zeromode}
\eea
(terms of still higher orders in $\eta$ produce
corrections small in the parameter $\Delta\tau_s\ll 1$), where
\begin{eqnarray}
&&{\bar k}_q(\br)=
V^{-1}\int  d^2\br' k_q(\br,\br'),
\nonumber \\
&&{\bar {\bar k}}_q=
V^{-2}\int  d^2\br d^2\br' k_q(\br,\br').
\label{kbar}
\end{eqnarray}

The contribution of non-zero modes
is found to be
\cite{BMM2}
\be
V^2\langle|\psi^2(\br_1) \psi^2(\br_2)|\rangle_{\rm
NZM}={\tilde \Pi}_B(\br_1,\br_2),
\label{tildeP}
\ee
where ${\tilde \Pi}_B(\br_1,\br_2)={\Pi}_B(\br_1,\br_2)-
{\Pi}^{(0)}_B(\br _1,\br_2)$ describes the (integrated over
direction of velocity) probability of classical propagation from
$\br_1$ to $\br_2$,
\begin{eqnarray}
&&\Pi_B(\br_1,\br_2)=\int\int d\bn_1
d\bn_2{\cal D}(\br_1\bn_1,\br_2\bn_2),\nonumber \\
&&{\hat{\cal L}}{\cal D}=(\pi\nu)^{-1}\left[\delta(\br_1-\br_2)
\delta(\bn_1-\bn_2)-V^{-1} \right],
\label{PB}
\end{eqnarray}
with the contribution ${\Pi}^0_B(\br _1,\br_2)$ of direct
propagation (before the first event of disorder scattering)
excluded. The symbol ${\hat{\cal L}}$ in (\ref{PB}) denotes the
Liouville
operator characterizing the classical motion \cite{footnote}.

We analyze now the total result given by the sum of
(\ref{C2zeromode}) and (\ref{tildeP}). First of all, we
stress that it satisfies exactly the constraint (\ref{norm}) of
wave function normalization.
Next, we consider sufficiently short distances,
$|\br_1-\br_2|\ll l_s$. In this case the correlation function is
dominated by the first term in the r.h.s. of
Eq.~(\ref{C2zeromode}), returning us to the result (\ref{ZM0}).
Furthermore, we can generalize this result to higher correlation
functions,
\bea
&&\langle\psi^*(\br_1)
\psi(\br'_1) \dots  \psi^*(\br_n)
\psi(\br'_n)\rangle =-\frac{1}{2V(n-1)!}\nonumber \\
&&\times\lim_{\eta\to 0}(2\pi\nu\eta)^{n-1}
\left\langle\sum_\sigma \prod_{i=1}^n{1 \over \pi \nu}
G_{p_i{p_{\sigma(i)}}}(\br_i,\br'_{\sigma(i)})
\right\rangle_{S[Q]}, \nonumber
\eea
where the summation goes over all
permutations $\sigma$ of the set $\{1,2,\dots,n\}$,
$p_i=1$ for $i=1,\dots,n-1$, and $p_n=2$.
If all the points $\br_i,\br_i^\prime$ are within a distance
$\ll l_s$ from each other, the leading contribution to this
correlation function is given by the zero-mode approximation with
higher-order terms in $\eta$ neglected [i.e. by the same
approximation which leads to Eq.~(\ref{ZM0})], yielding
\bea
&&V^n\langle\psi^*(\br_1)
\psi(\br'_1) \dots  \psi^*(\br_n)
\psi(\br'_n)\rangle=\sum_\sigma\prod_{i=1}^{n}
f_F(\br_i,\br_{\sigma(i)}^\prime),\nonumber \\
&&f_F(\br,\br')=-{\rm Im}G^R(\br,\br')/(\pi\nu).
\label{fF}
\eea
This result
is identical to the statement
(\ref{Ppsi}) of the Gaussian statistics of eigenfunctions.
We have thus proven that the conjecture of Refs. \cite{Berry77},
\cite{Srednicki2} holds within a spatial region of an extension
$\ll l_s$, with the kernel $C(\br_1,\br_2)$ given by
Eqs.~(\ref{fF}),\ (\ref{GRA}).

We turn now to the behavior of the correlator
$\langle|\psi^2(\br_1) \psi^2(\br_2)|\rangle$
at larger separations $|\br_1-\br_2|\gg l_s$.
In this situation, the correlations are dominated by the
contribution (\ref{tildeP}) of non-zero modes. Let us further
note that the smooth part of the zero-mode contribution
(\ref{C2zeromode}) (i.e. with Friedel-type oscillations on the
scale of the wave length $\lambda=2\pi/k$ suppressed) is
exactly equal to ${\Pi}^{(0)}_B$. Therefore, the smoothed
correlation function is given by the classical propagator,
\be
V^2\langle|\psi^2(\br_1) \psi^2(\br_2)|\rangle_{\rm smooth}-1=
\Pi_B(\br_1,\br_2),
\label{smooth}
\ee
independent of the relation between $|\br_1-\br_2|$ and $ l_s$.
The mean free path $l_s$ manifests itself only in setting the
scale on which the oscillatory part of
$\langle|\psi^2(\br_1) \psi^2(\br_2)|\rangle$ gets damped.

A result for the variance of matrix
elements related to Eq.~(\ref{smooth}) was obtained in
\cite{Eckhardt95} by a semiclassical method. Note, however, that
the semiclassical treatment of \cite{Eckhardt95} is only
justified if one introduces a sufficiently large level broadening
$\eta \gg \Delta$, while calculating
statistical properties of a single eigenfunction requires the
limit $\eta \ll \Delta$, see Eq.~(\ref{C2fG}). Also, the result
of \cite{Eckhardt95} does not contain the contribution
corresponding to the term $V^{-1}$ in (\ref{PB}), so that it
violates the wave function normalization
in the same way as the conjecture (\ref{Ppsi}),\ (\ref{sred}).

Since we have shown that for $l_s\ll L$ the applicability of the
Gaussian statistics (\ref{Ppsi}),\ (\ref{sred}) is restricted
to the scales $\ll l_s$, one may be tempted to ask whether
increasing $l_s$ beyond $L$ would
be favorable from this
point of view. The answer is no; in contrast, for $l_s \gtrsim L$
a further increase of $l_s$ reduces the range of applicability of
the Gaussian statistics. Indeed, it is not difficult to show that
for $l_s \gg L$ the Green's function (\ref{GRA}) has the form
$G^R(\br_1,\br_2)\simeq J_0(k|\br_1-\br_2|)$ (we assume for
simplicity that the points $\br_1,\br_2$ are sufficiently far
from the boundary) only for $|\br_1-\br_2|\ll {\tilde l}_s$,
where ${\tilde l}_s=L^2/l_s$. At larger distances,
$|\br_1-\br_2|\gtrsim {\tilde l}_s$, the Green's function shows
irregular oscillations with a characteristic amplitude
$|G^R(\br_1,\br_2)| \sim (k{\tilde l}_s)^{-1/2}$ independent of
$|\br_1-\br_2|$, which are physically due to the interference
of waves multiply reflected from the boundary.
Therefore, only at $|\br_1-\br_2|\ll {\tilde l}_s$ the first term
in (\ref{C2zeromode}) will dominate and the statistics will be
Gaussian.

{\it Random magnetic field.}
In the above we studied the wave function statistics of a given
chaotic system by generating an ensemble of quantum systems with
the help of an additional random potential. Now we use the same
approach to study the wave function statistics in a random
magnetic field (RMF). In this case, an ensemble is
defined from the very beginning and the introduction of
additional weak disorder may be considered as a technical trick,
the reason for which is explained below.

We consider a white-noise RMF $B(\br)$ with the correlation
function
\begin{equation}
\langle B(\br)B(\br')\rangle = \Gamma\delta^{(2)}(\br-\br'),
\qquad \Gamma\ll k^2,
\label{HH}
\end{equation}
and assume that the size of the system, $L$, is sufficiently
large. On length scales longer than the transport mean free path
$l_{tr}=4k/\Gamma$ this problem is described by the
conventional unitary-class diffusive $\sigma$-model
\cite{amw} so that the results obtained for diffusive
systems \cite{PhysRep} apply. We will be interested, however,
in wave function correlations on much shorter -- ballistic --
length scales. Specifically, we will study how the Friedel-type
oscillations in
$\langle |\psi^2(\br_1)\psi^2(\br_2)| \rangle$
decay with increasing $|\br_1-\br_2|$. [The smooth part is simply
$V^2\langle |\psi^2(\br_1)\psi^2(\br_2)| \rangle=(\pi k
|\br_1-\br_2|)^{-1}$ for all $|\br_1-\br_2|\ll l_{tr},\: L,$
as follows from (\ref{smooth})].

In the case of a random potential the scale for the vanishing of
oscillations is set by the single-particle mean free path $l_s$.
An attempt to
get
an analog of this result by deriving
directly the ballistic $\sigma$-model via averaging over the RMF
fails, since the equation for $l_s$ obtained within the
self-consistent Born approximation (SCBA) leads to an
infrared-divergent and gauge-dependent
result \cite{AAMW}. This is a manifestation of the fact that in
the case of a RMF the single-particle relaxation rate depends on
geometry of the problem.

To overcome this problem, we add an additional weak random
potential with the mean free path $l_s^{\rm RP}$ much longer than
the length scale of interest set by the RMF (which we will find
below). Averaging over this random potential, we derive the
$\sigma$-model in a given realization of the RMF. As explained
above, the two-point correlation function of eigenfunction
intensities on a scale $|\br_1-\br_2|\ll l_s^{\rm RP}$
is given by Eqs.~(\ref{ZM0}),\ (\ref{kq}). Therefore, the desired
oscillatory contribution reads
\begin{eqnarray}
&&\langle k_q^{\rm osc}(\br_1,\br_2) \rangle
_{\rm RMF}\nonumber \\
&&=-(\pi\nu)^{-2}
{\rm Re}\left\langle G^R(\br_1,\br_2)G^R(\br_2,\br_1)
\right \rangle
_{\rm RMF},
\label{kqRMF}
\end{eqnarray}
where $G^R=(E-{\hat H} + i/2\tau_s^{\rm RP})^{-1}$
is the Green's function in a given realization of the RMF, and
$\langle \dots \rangle_{\rm RMF}$ denotes averaging over the RMF
realizations. This (second) averaging can be performed with use
of the path integral formalism \cite{MAW}.
The product of the two Green's functions in (\ref{kqRMF})
can be written as
\begin{eqnarray}
&&\langle G^R({\bf R},0)G^R(0,{\bf R})\rangle_{\rm RMF}
=\int_0^\infty \! \!\! dT_1 dT_2
\int_{\br_i(0)=0}^{\br_i(T_i)={\bf R}}\! \! \! \! {\cal D}\br_1
{\cal D}\br_2 \nonumber \\
&&\times \exp[i(E+i/2\tau_s^{\rm RP})(T_1+T_2)+iS_{\rm
kin}-S_{\rm RMF}],
\label{GG}
\end{eqnarray}
where $S_{\rm kin}=\int_0^{T_1}dt \ m{\dot
\br}_1^2/2+\int_0^{T_2}dt \ m{\dot \br}_2^2/2$, and $S_{\rm
RMF}=\Gamma s_{\rm no}/2,$ with $s_{\rm no}$ denoting the
non-oriented area between the two trajectories $\br_1(t)$ and
$\br_2(t)$. The integral (\ref{GG}) is dominated by the pairs of
paths being close to each other and corresponding to an almost
uniform and straight motion from $0$ to ${\bf R}$. To make this
explicit, it is useful to perform the change
of variables \cite{MAW}, introducing $\br=\br_1-\br_2, \ {\bbox
\rho}=(\br_1+\br_2)/2, \ t_{+}=(T_1+T_2)/2$, and $t_{-}=T_2-T_1
\; (t_{+}\gg t_{-})$. The RMF-induced part of the action takes
then the form
\be
S_{\rm RMF}=\frac{v\Gamma}{2}\int_0^{t_{+}} dt|r^{\perp}(t)|,
\label{SRMF}
\ee
where $v=k/m$ is the particle velocity, and we have split $\br$
into components parallel ($r^{||}$) and
perpendicular
($r^{\perp}$) to ${\bbox {\dot \rho}} \approx {\bf R}/t_{+}$.
While the integrals over ${\bbox \rho}$ and $r^{||}$
are essentially the same as for a free particle, that
over $r^{\perp}$ has the form of the Feynman integral
for a one-dimensional
particle in the potential $i(v\Gamma/2)|x|$.
The corresponding Green's function $g(x,x',t)$
reads in the frequency
representation at $x=x'=0$ (which is what we need in view of the
boundary conditions on $r^{\perp}$),
\begin{equation}
g(0,0,\omega)=-\frac{i^{-1/3}}{2}(m\tau_{0})^{1/2}
\frac{{\rm Ai}(-i^{-2/3}\omega\tau_{0})}
{{\rm Ai}'(-i^{-2/3}\omega\tau_{0})},
\label{g00}
\end{equation}
where ${\rm Ai}(z)$ is the Airy function, and
\begin{equation}
\tau_{0}=(4m/\Gamma^2 v^2)^{1/3}.
\label{tauo}
\end{equation}
This leads to the following result for the oscillatory part of
the wave function correlation function
\bea
&&V^2\langle|\psi^2(\br_1) \psi^2(\br_2)|\rangle^{\rm osc}_{\rm
RMF} =\frac{1}{\pi k r}\nonumber \\
&&\times \left\{
\begin{array}{ll}
\sin(2k r), & \!  r\ll l_{\rm 0}\vspace{0.1cm}\\
\sin\left(2k r+|\zeta_0|r/2l_{0}-\pi/{12} \right)\\
\times \left(\pi r/l_{0}\right)^{1/2}
\exp\left[-\sqrt{3}|\zeta_0 | r/2l_{0}\right] \!, &
r\gg l_{0} \end{array}
\right.
\label{RMFcorr}
\eea
where $r=|\br_1-\br_2|$,
$\zeta_0\simeq -1.05$ is the lowest zero of ${\rm Ai}'(z)$,
and $l_{0}=v\tau_{0}$.

We thus find that the oscillations are suppressed on the
scale $\sim l_{0}=v\tau_{0}=(4k/\Gamma^2)^{1/3}$.
Note that $l_{0}$ is parametrically different from both
the transport mean free path $l_{tr}=4k/\Gamma$ and the
length $l_{\rm dHvA}=v\tau_{\rm dHvA}=(2\pi/\Gamma)^{1/2}$
characterizing damping of de Haas - van Alphen
magnetooscillations of the density of states,
$\rho_{\rm osc} \propto \exp[-(\pi/\omega_c\tau_{\rm dHvA})^2]$
\cite{MAW}.
Difference between $l_{0}$ and $l_{\rm dHvA}$ (in
the case of a random potential both these scales are
set by $l_s$) illustrates the already mentioned dependence of the
single-particle relaxation rate on the geometry of the problem
in the case of RMF.

The scale $l_{0}$ was obtained in
\cite{AI} from
consideration of certain Green's function with an obscure
physical meaning. We have demonstrated that the length
$l_{0}$ determines an observable quantity -- the
scale of decay of the oscillatory part of the wave function
correlation function.

{\it Conclusions.} We have studied the wave function
statistics in a chaotic ballistic system. The corresponding
statistical ensemble is defined by adding a smooth random
potential, satisfying $l_{tr}\gg L \gg l_s$. The first
inequality guarantees that the random potential does not change
the classical dynamics, while the second one ensures that the
ensemble of quantum systems is sufficiently large and provides
meaningful statistics. By using the ballistic $\sigma$-model
approach we have proven that the conjecture of Gaussian
fluctuations of wave functions \cite{Berry77,Srednicki2}
holds on sufficiently short distances $|\br_i-\br_j| \ll l_s$,
while it is strongly violated on larger scales. Our results
solve, in particular, the problem of inconsistency of the
conjecture of Gaussian statistics with the wave function
normalization.

We have further applied these results to study the decay
of Friedel-type oscillations in the
correlation function
$\langle|\psi^2(\br_1) \psi^2(\br_2)|\rangle$ in a RMF.
In this case averaging over an additional weak random
potential yields Gaussian fluctuations of wave functions in a
given realization of the RMF.
The remaining averaging over the RMF realizations
performed via the path integral formalism leads to the result
(\ref{RMFcorr}). The scale $l_{0}$ for the decay of
oscillations (playing the role of the single-particle mean free
path $l_s$) is given by Eq.~(\ref{tauo}), providing physical
meaning to a length found in \cite{AI} from some formal
consideration.

We thank P.~W\"olfle for useful comments.
This work was supported by the SFB195
and the Schwerpunktprogramm
``Quanten-Hall-Systeme'' der Deutschen
Forschungsgemeinschaft, the INTAS grant 99-1705,
and the RFBR grant 99-02-17093.

\end{multicols}


\begin{references}
\bibitem{footnote} In fact, the operator ${\hat{\cal L}}$ in
Eq.~(\ref{PB}) contains an additional term describing scattering
by the random potential, $\delta{\hat{\cal L}}f(\bn)=\int
d\bn'[f(\bn)-f(\bn')]w(\bn,\bn')$. However, this term does not
affect essentially the propagator $\Pi_B(\br_1,\br_2)$ in view
of the assumption $l_{tr}\gg L$.
\end{references}
\end{document}